\documentclass[journal]{IEEEtran}

\usepackage{graphicx}
\usepackage{url}
\usepackage[dvipsnames]{xcolor}
\usepackage{makecell}
\usepackage{amsfonts}
\usepackage[normalem]{ulem}

\ifCLASSINFOpdf
\else
\fi
\begin{document}
%
\title{Some comments about CRC selection for the 5G NR specification}
%
%
%

\author{Tsonka~Baicheva,~\IEEEmembership{Member,~IEEE,}
        Peter~Kazakov,
        and~Miroslav~Dimitrov,~\IEEEmembership{Student~Member,~IEEE}
\IEEEcompsocitemizethanks{\IEEEcompsocthanksitem T. Baicheva, P. Kazakov and M. Dimitrov were with the Institute of Mathematics and Informatics, Bulgarian Academy of Sciences, Sofia, Bulgaria. P. Kazakov was also with AtScale, Sofia, Bulgaria. }}

%
%

\markboth{Journal of \LaTeX\ Class Files,~Vol.~14, No.~8, August~2015}%
{Shell \MakeLowercase{\textit{et al.}}: Bare Demo of IEEEtran.cls for IEEE Journals}
%



\maketitle

\begin{abstract}
In this work the undetected error probability performance of the proposed in the technical specification of 5G new radio cyclic redundancy-check codes taking into consideration that they have to work for fixed interval of lengths are investigated. Codes of 6, 11 and 16 check bits with better error detection performance than the proposed in the specification ones are suggested. For 24 check bits the 3 best CRC codes are presented, but no only one best for the whole interval of lengths code was found.
\end{abstract}


%
\IEEEpeerreviewmaketitle

\section{Introduction}
\label{sec:introduction}

According to the technical specification of the 5G new radio (5G NR) \cite{3gpp}, data and control streams from/to MAC layer are encoded/decoded to offer transport and control services over the radio transmission link. Channel coding scheme is a combination of error detection, error correction, rate matching, interleaving and transport channel or control information mapping onto/splitting from physical channels. For error detection cyclic redundancy-check (CRC) codes or combination of LDPC and CRC codes are used. Six CRC codes are proposed in the specification and they are specified to work for different payloads (information stream lengths) $A$. The proposed CRC codes are summarized in Table \ref{tab:CRCcodes}.

\renewcommand{\arraystretch}{1.8}

\begin{table}[h!]
\caption{CRC codes used in 5G NR}
\label{tab:CRCcodes}
\begin{center}
\begin{tabular}{|l|c|}
\hline
Label & Polynomial\\
\hline
CRC6 & $x^6+x^5+1$ \\
\hline
CRC11 & $x^{11}+x^{10}+x^9+x^5+1$ \\
\hline
CRC16     & $x^{16}+x^{12}+x^5+1$\\
\hline
CRC24A     & \makecell{$x^{24}+x^{23}+x^{18}+x^{17}++x^{14}+$ \\ $+x^{11}+x^{10}+x^7+x^6+x^5+x^4+x^3+x+1$}\\
\hline
CRC24B & $x^{24}+x^{23}+x^6+x^5+x+1$ \\
\hline
CRC24C & \makecell{$x^{24}+x^{23}+x^{21}+x^{20}+x^{17}+x^{15}+$ \\ $+x^{13}+x^{12}+x^8+x^4+x^2+x+1$} \\
\hline
\end{tabular}
\end{center}
\end{table}

For the uplink and downlink shared channels, CRC16 should be used if the transport block size $A$ is $1\leq A\leq 3824$, while for $3824\le A\leq 8424$, CRC24A is proposed. LDPC codes are combined with CRC24B for block sizes $A\leq 8424$. For the downlink broadcast channel, where $A\leq 8424$, CRC24C is suggested. Error detection in the uplink control channel for $12\le A\leq 19$ should be done by CRC6, for $20\le A\leq 1706$ by CRC11 while in the downlink control channel, where  $A\leq 140$, CRC24C should be used. 

In this letter we investigate the error detection performance of the aforementioned CRC codes. The work is organized as follows. First we briefly recall the basic techniques of measuring the error detection performance of a given CRC code. Then we suggest different collection of six CRC codes which outperform, in terms of undetected error probability, those proposed in 5G NR. We compare the two collections of codes by applying both the optimization criterion and the additionally speed up investigation method suggested in our previous work \cite{BK}.  Although that in \cite{BK} we obtain best CRC codes for 11 and 16 check bits we did the investigation again as in the 5G NR specification CRC codes have to work in a fixed range of lengths which do not coincide with those in \cite{BK}. As the results show, the best CRC code with 11 check bits is different than the obtained in \cite{BK}.

\section{Basic facts about CRC codes}

CRC codes are first introduced by Peterson and Brown in \cite{PB}. Very good sources describing the theoretical foundations and properties of CRC codes are Peterson and Weldon's \cite{PW} and Wicker's \cite{W} books. In this section, the main results regarding the structure and error detection properties of CRC codes are summarized and introduced.

CRC codes are full length or shortened linear binary cyclic codes. Each codeword of a CRC code has $n=k+p$ binary digits and is obtained by adding in a definite way a block 
of $p$ parity bits to an information block of $k$ binary digits 

 Any CRC code may be represented as a set of polynomials where instead of the information block of $k$ binary digits, the polynomial $$i(x)=i_0+i_1x+\dots+i_{k-1}x^{k-1},$$ and instead of the block of $p$ parity bits the polynomial $$r(x)=r_0+r_1x+\dots+r_{p-1}x^{p-1} $$ are used. Then in each CRC code with $p$ check bits there exists a polynomial of degree $p$ called the {\it generator polynomial} $g(x)$ of the code. The leading and zero coefficients of $g(x)$ are nonzero.

 Each codeword $c(x)$ of the CRC code is obtained by computing $r(x)$ from $i(x)$ in the following way $$r(x)\equiv (x^p.i(x))~mod~g(x).$$ According to this encoding the following equations holds $$c(x)=x^pi(x)+r(x)=q(x)g(x)+r(x)+r(x)=q(x)g(x)$$ for some polynomial $q(x)$. Thus, the recipient could easily decides if a given word is a codeword or not by checking its divisibility to $g(x)$. 

The polynomial $g(x)$ divides $x^{n_c}+1$, where $$n_c=\min \{m\vert x^m\equiv 1~mod~g(x)\}$$ is called {\it order of the polynomial} $g(x)$. The binary cyclic code $D$ generated by the polynomial $g(x)$ of degree $p$ is a $[n_c,n_c-p]$ code. Any $[n,n-p]$ subcode $C$ of $D$ obtained by shortening $D$ in $n_c-n$ positions is a CRC code. The maximum length at which any CRC code can be used is equal to the order of the generator polynomial $n_c$. For longer lengths the code is no more CRC code. More precisely, it becomes a repetition code with minimum distance 2. This means that it can detect only 1 error and does not posses the ability to detect burst errors.  

Given their excellent error detection capabilities, CRC codes are widely exploited by manifold applications.

In summary, CRC codes of length $n$ and minimum distance $d$, generated by the polynomial $g(x)$ of degree $p$, could detect:

\begin{itemize}
\item{all combinations of up to $d-1$ errors;}
\item{all error bursts (any pattern of errors for which the number of symbols between the first and last errors, including the errors, is $b$) of length $b\leq p$;}
\item{a fraction of all bursts of length $b=p+1$ (the fraction equals $1-2^{-(p-1)}$);}
\end{itemize}

Let us consider communication through a Binary Symmetric Channel (BSC). BSC is a discrete memoryless channel with binary input and output and crossover probability $\varepsilon$, where $0\leq \varepsilon\leq 1/2$. During the transmission, the send codeword $c(x)$ may be changed by the channel noise and the receiver to get another polynomial $c'(x)=c(x)+e(x)$. The nonzero coefficients of the error polynomial $e(x)$ correspond to the positions which are changed by the noise. At the receiving end $c'(x)$ is divided by the generator polynomial $g(x)$ and an error is detected if the reminder of the division is not zero. However, it may happened that the channel noise changes $c(x)$ in such a way that $c'(x)$ is another codeword. In this case we have an {\it undetected error}. As the CRC codes are linear, we will have undetected error if and only if $e(x)$ is a codeword. 

Let us denote by $A_i$ and $B_i$ the number of codewords of weight $i$ in the CRC code and in its dual. Then the error detection performance of a CRC code of length $n$ with $p$ check symbols is measured by the probability $P_{ue}$ of undetectable errors \cite{WML}:
$$ P_{ue}=\sum_{i=1}^n A_i\varepsilon^i(1-\varepsilon)^{n-i}=2^{-p}[1+\sum_{i=1}^n B_i(1-2\varepsilon)^i]-(1-\varepsilon)^n.$$
Therefore, the best choose of a CRC code is the one having the smallest value of $P_{ue}$. As $P_{ue}$ depends on the crossover probability $\varepsilon$ and the length $n$ of the CRC code we may have different best codes for different $\varepsilon$ and $n$.

\section{Complexity of the determination of $P_{ue}$ of a code} 

Let us denote by $d$ the minimum distance of an $[n,n-p]$ CRC code generated by the polynomial $g(x)$ and shortened in $n_c-n$ positions. 
It has been proven in \cite{V} that determination of the set $\{A_i\}_{i=1}^n$ (or $\{B_i\}_{i=1}^n$ respectively) for a $[n,n-p]$ code is a computationally hard problem. Thus the comparison of the values of $P_{ue}$ for two particular codes is also computationally hard problem. 

In order to reduce the computational complexity of the problem we could compare $P_{ue}'=A_d\varepsilon^d(1-\varepsilon)^{n-d}$, i.e. to calculate only the first nonzero addend of the sum for $P_{ue}$, as it is the most significant one (especially for small values of $\varepsilon$). By using this approach, only the number of codewords with minimum weight is needed.

\begin{figure}
\centering
\caption{Comparison of $P_{ue}$ and $P_{ue}'$ of two CRC codes}
\label{PueComp}
\includegraphics[width=1\linewidth]{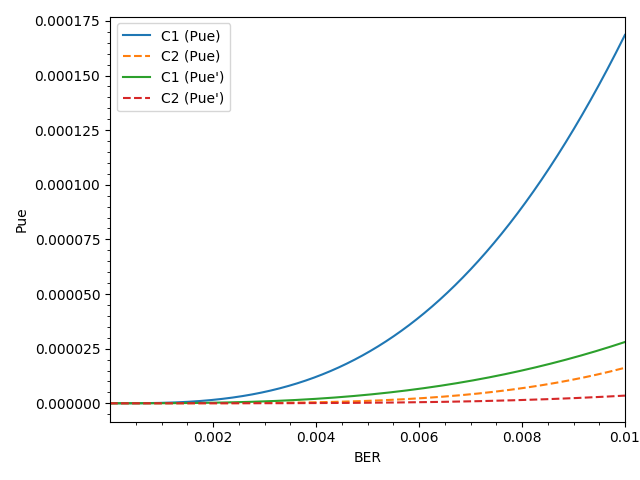}
\end{figure}

Comparing $P_{ue}'$ performances of two codes we can obtain rather precise result as it can be seen in Figure \ref{PueComp}. The actual difference between the
the undetected error probabilities of the compared codes is several times bigger than that obtained by comparing only its $P_{ue}'$s. This approach has been used by Kazakov in \cite{Kaz} to determine optimal with respect to $A_d$ CRC codes with 16 check bits.  

\section{Search for optimal CRC codes with $p$ check symbols}

The proposed in the 5G NR specification CRC codes will be used to protect information packages of different lengths. That is way our aim is to chose a CRC code having the best performance not only for a particular length but for the whole range of lengths at which the code is supposed to work. 

We denote by $d(C_{g, n})$ the minimum distance of a CRC code of length $n$, with generator polynomial $g(x)$, supposed to work in an interval of lengths between $L$ and $M$. Since the CRC code can be used at any length of the interval $[L\dots M]$ its cumulative performance in the whole interval should be investigated. Thus, for any polynomial $g(x)$ of degree $p$, which could be generator polynomial of a CRC code of length $n \in [L\dots M]]$ we calculate

$$S_d = \sum_{n=L}^{M}d(C_{g, n}).$$ 
The CRC code with a maximum value of $S_d$ is the optimal one. 

In order to obtain $d(C_{g, n})$, we have to generate all $2^{n-p}$ codewords of the CRC code and to determine their weights as the minimum distance of a binary linear code is equal to the minimum weight of its codewords. Given some fixed CRC code, to calculate $S_d$ we have to repeat this procedure $M-L+1$ times. However, instead to generate the $2^{n-p}$ codewords of the CRC code $C$, we can generate only the $2^p$ codewords of its dual code $C'$ and compute the minimum distance of $C$ in linear time via MacWilliams' identities \cite{MacW}.

To calculate the values of $S_d$ for all CRC codes with $p=6,11,16,24$ parity bits we apply the same divide and conquer technique suggested in \cite{BK}. To speed up the algorithm additionally, we use a new early rejection criteria by requiring minimum distance 6 for large code lengths and apply some improved programming techniques. Then for each $p=6,11,16,24$ we determine the CRC code having maximum $S_d$. If we obtain more than one code of maximum $S_d$, we  consider those having the smallest sum of the numbers of codewords with minimum weight, i.e. minimum $S_{A_d}=\sum_{n=L}^{M}A_d$.

\section{Results}

In Table \ref{CRClengths} we summarize the range of code lengths at which the proposed in 5G NR technical specification CRC codes are defined to work.

\begin{table}[h!]
\caption{Lengths at which CRC codes from 5G NR specification are applicable.}
\label{CRClengths}
\begin{center}
\begin{tabular}{|c|c|c|}
\hline
CRC code &$L$ & $M$\\
\hline
CRC6   & 18 & 25	\\
CRC11  & 31 & 1717 \\
CRC16  & 17 & 3840 \\
CRC24A & 3848 & 8448 \\
CRC24B & 25 & 8448 \\
CRC24C & 25 & 164 or 8448  \\
\hline
\end{tabular}
\end{center}
\end{table}

For CRC codes with $p=6,11,16$ or 24 parity bits, we determine the best ones, by following the strategy described throughout the previous section. We consider only CRC codes of lengths in the interval $[L\dots M]$ according to Table \ref{CRClengths}. Given a fixed number of parity bits $p$, we have to consider $2^{p-1}$ polynomials - possible generators of a CRC code. However, we could reduce this number by omitting the reciprocal ones, since they generate equivalent codes or codes with orders smaller than $M$. 

\subsection{CRC codes of $p=6$}

All polynomials of degree $6$ that could generate CRC codes are $2^5=32$. After excluding the reciprocal they remain 19. Only 8 polynomials out of these 19 have orders greater than 25 (for $p=6$, $M=25$) and we consider only them. In Table \ref{T-CRC6} we summarize the obtained results. The generator polynomials of the CRC codes are presented in a hexadecimal notation, i.e. $x^6+x^5+x^4+x+1$ is $73$, while their respective reciprocal polynomials are given in the brackets. The weight spectra of the first two codes are not calculated, because they have smaller $S_d$ than the others. In the table are also provided the order $n_c$ of the polynomial, the minimum distances of the CRC codes and the interval of lengths for which the code has the corresponding minimum distance.

\begin{table}[h!]
\caption{Results for CRC codes with 6 check symbols.}
\label{T-CRC6}
\begin{center}
\begin{tabular}{|l|c|c|c|c|}
\hline
$g(x)$ &$n_c$ & $S_d$ & $S_{A_d}$& minimum distances\\
\hline
73 (67)     & 63 & 24 & - & -\\
6d (5b)     & 63 & 24 & - & - \\
61 (43) CRC6& 63 & 24 & 173 & 3,18...25\\
47 (71)     & 31 & 32 & 1959 & 4,18...25\\
59 (4d) optimal  & 31 & 32 & 1956 & 4,18...25\\
7b (6f)     & 31 & 32 & 1966 & 4,18...25 \\
7d (5f)     & 30 & 32 & 1962 & 4,18...25\\
4b (96)     & 28 & 32 & 1959 & 4,18...25 \\
\hline
\end{tabular}
\end{center}
\end{table}

\begin{figure}[h]
\caption{Comparison of $P_{ue}'$ of CRC6 and the optimal CRC codes.}
\label{CRC6}
\centering
\includegraphics[width=1\linewidth]{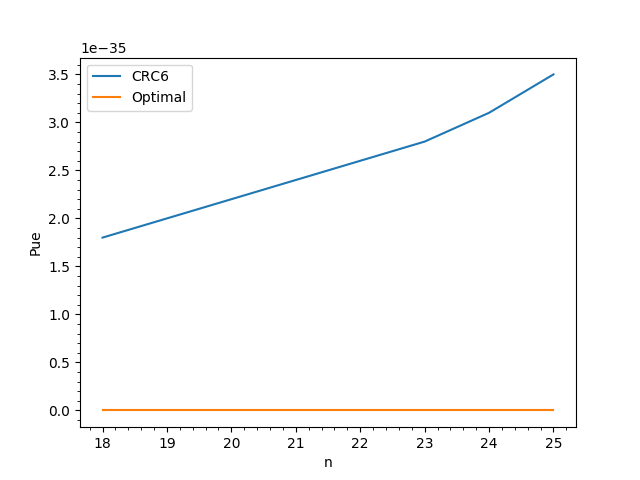}
\end{figure}

According to our optimization criterion, the CRC code generated by the polynomial 59 is optimal. However, all five codes with a maximum value of $S_d$ have very close values of $S_{A_d}$. Thus the differences of their $P_{ue}$ values are negligible and any of them could be a good choice. The code with generator polynomial 47 is also suggested by Philip Koopman in its "Best CRC polynomials" web cite \cite{K} as the best general-purpose polynomial generating CRC code with 6 check bits (as 0x23 with HD=4) and having maximum minimum distance profile. The CRC code with generator polynomial 61 is the proposed in 5G NR CRC6 code and any of the five codes with $S_d=32$ will have much better performance than them. 
In Figure \ref{CRC6} we present comparison between the $P_{ue}'$ of the CRC codes with generator polynomials 61 and 59 for $\varepsilon=10^{-12}$ and all lengths $n$ at which the code is supposed to be used. The suggested by us optimal code is at least 100\% better than CRC6 at all lengths.  This result is not surprising as CRC6 has minimum distance 3 for all lengths from 18 to 25 while the optimal CRC code has minimum distance 4.

\subsection{CRC codes of $p=11$}

There are $2^{10}=1024$ possible choices for generator polynomials of CRC codes of degree $11$. After excluding the reciprocal and those of order smaller than 1717, we obtain that the CRC code generated by the polynomial e0f has maximum $S_d$. In Table \ref{T-CRC11} we present the results regarding the CRC codes generated by the e0f polynomial and the proposed in 5G NR one generated by the polynomial e21 (CRC11). The proposed in \cite{BK} 93f polynomial is not optimal in this investigation as it has maximum $S_d$ for lengths between 12 and 512 but not for the longer interval of lengths between between 31 and 1717.  The CRC code generated by the e0f polynomial possesses significantly better performance than CRC11 in the interval of lengths $55\dots 149$ where it has minimum distance 4 while CRC11 has 3. For example, for $\varepsilon=10^{-12}$ at length 130 the optimal code is 100\% better than CRC11, at length 300 it is 50\% better, while at length 1000 it is 0,3\% better.   

\begin{table}[h!]
\caption{Results for CRC codes with 11 check symbols.}
\label{T-CRC11}
\begin{center}
\begin{tabular}{|l|c|c|c|}
\hline
$g(x)$ &$n_c$ & $S_d$ &minimum distances\\
\hline
e0f (f07) & 1953 & 5180 &3,150...1717;4,31...149\\
e21 (847) CRC11& 2047 & 5085 &3,55...1717;4,31...53\\
\hline
\end{tabular}
\end{center}
\end{table}

\subsection{CRC codes of $p=16$}

For CRC codes with 16 check bits, the maximum $S_d$ for lengths up to 3840 has the polynomial 1a2eb. In Table \ref{T-CRC16} the values of $S_d$ for the polynomials 1a2eb and 11021 (CRC16) are given. The polynomial 1a2eb is suggested by Funk \cite{F} (321353 in his notation) as the polynomial which generates CRC codes with the best undetected error probability performance for the whole range of lengths (17..32767) it can be used. This polynomial is also suggested by Koopman \cite{K} in the maintained by him website as well as in our previous work \cite{BK}. Its performance will be much more better for lengths up to 109. For example, for $\varepsilon=10^{-12}$ at lengths 23 and 100 the optimal code is 100\% better than CRC16, at length 620 it is 2\% better and at length 2500 it is 0,8\% better.   

\begin{table}[h!]
\caption{Results for CRC codes with 16 check symbols.}
\label{T-CRC16}
\begin{center}
\begin{tabular}{|l|c|c|c|}
\hline
$g(x)$ &$n_c$ & $S_d$&minimum distances\\
\hline
1a2eb (1ae8b) & 32767 & 15508 &4,110...3840;6,28...109;\\
& & & 8,19...27;10,17...18\\
\hline
11021 (10811) CRC16 & 32767 & 15296 & 4,17...3840\\
\hline
\end{tabular}
\end{center}
\end{table}

\subsection{CRC codes of $p=24$}

Since all of the three suggested CRC codes with 24 check bits should work for lengths up to 8448 we put $M=8448$. The results about the first three best polynomials obtained by us, together with those three suggested in 5G NR, are given in Table \ref{T-CRC24}. In Table \ref{D-CRC24} the minimum distances of the codes from Table \ref{T-CRC24} are given. The suggested CRC24A and CRC24B polynomials are standardized ones known as CRC-24-Radix-64 and WCDMA 24 correspondingly. The three suggested in 5G NR CRC24 polynomials are good in general and can be found on Koopman's website \cite{K}. In our investigation we take into consideration the interval of lengths (i.e. $[L\dots M]$) for which the codes will be used and find the best for this particular interval of lengths codes. Thus the three codes suggested by us have the best values of $S_d$ for the interval $[25\dots 8448]$. In contrast with the CRC codes of 6, 11 and 16 bits where there is unique code which has the best performance in the whole interval $[L\dots M]$ for codes with 24 bits different codes are optimal for different lengths and we present the three best ones according to our optimization criterion. For example, at length 920 the three optimal codes are between 94\% and 100\% better than the three suggested in 5G NR CRC24 codes. For the code lengths from the interval $[3848\dots 8448]$ CRC24A, CRC24B and the suggested by us three codes have minimum distance 4 (see Table \ref{D-CRC24}) for the whole interval and their performance is very close. For example, 
the optimal codes are only 2\% better than CRC24A at length 4000 and they perform slightly worse for lengths above 7060 where CRCA has smaller number of codewords of minimum weight. As the suggested by us three codes have better values of $S_d$ they will perform better in general (i.e. when they are used at different lengths) ant their slightly worst performance at lengths close to the end of the interval $[3848\dots 8448]$ will be compensated by their much better performance at shorter lengths.  

The suggested in \cite{BK} 11175b7 polynomial generates CRC codes with best value of $S_d$ for lengths between 25 and 512. It has order 1195740 and could be used in the interval $[25\dots 8448]$ but would has worse performance than the suggested in this work three CRC codes because it has minimum distance 6 only to length 586. This example illustrates once again the necessity to take into consideration the particular conditions (length, bit error rate) at which the code will be used in order to obtain the best one. 

We do not consider the interval $[25\dots 164]$ since the same code will also be used for the interval $[25\dots 8448]$. We only note for a completeness that for the interval $[25\dots 164]$ CRC24C has better performance with $S_d=924$ than the suggested by us codes where the best $S_d=802$ is achieved by 118b983. Thus CRC24C is a good choice if it is used only for lengths between 25 and 164. 

\begin{table}[h!]
\caption{Results for CRC codes with 24 check symbols.}
\label{T-CRC24}
\begin{center}
\begin{tabular}{|l|c|c|}
\hline
$g(x)$ &$n_c$ & $S_d$ \\
\hline
118b933 (1993a31)       & 139230  & 35584 \\
125ae5d (174eb49)       & 6241542 & 35564\\
10f6f6d (16dede1)       & 294903  & 35548\\
1864cfb (1be64c3) CRC24A& 8388607 & 34816\\
1800063 (18c0003) CRC24B& 8388607 & 33704\\
1b2b117 (1d11a9b) CRC24C& 28062   & 31109\\
\hline
\end{tabular}
\end{center}
\end{table}

\begin{table}[h!]
\caption{Minimum distances for CRC codes with 24 check symbols.}
\label{D-CRC24}
\begin{center}
\begin{tabular}{|l|c|}
\hline
$g(x)$ &minimum distances  \\
\hline
118b933 (1993a31)       & 4,916...8448; 6,60...915;8,39...59;\\
&10,29...38;12,25...28 \\ 
\hline
125ae5d (174eb49)       & 4,896...8448; 6,64...895;8,43...63;\\
&10,28...42;14,25...27\\
\hline
10f6f6d (16dede1)       & 4,881...8448; 6,74...880;8,42...73;\\
&10,27...41;14,25...26\\
\hline
1864cfb (1be64c3) CRC24A& 4,542...8448; 6,55...541;8,34...54;\\
&10,28...33;12,26...27;14,25\\ 
\hline
1800063 (18c0003) CRC24B& 4,29...8448; 6,25...28\\
\hline
1b2b117 (1d11a9b) CRC24C& 3,5135...8448; 4,504...5134;5,503...182;\\
&6,52...181;7,43...51;9,32...42;13,25...31\\
\hline
\end{tabular}
\end{center}
\end{table}

\section{Conclusion}
The error control performance of CRC codes with 6, 11, 16 and 24 check bits for the particular lengths at which the proposed in 
the technical specification of 5G new radio CRC codes are supposed to be used is investigated. For each number of check bits optimal codes are suggested and their performance is compared to the performance of the aforementioned proposed codes.

\ifCLASSOPTIONcompsoc
  \section*{Acknowledgments}
\else
  \section*{Acknowledgment}
\fi

The work of the first and the third authors has been partially supported by the Bulgarian National Science Fund under contract number 12/8 15.12.2017. The work of the second author has been partially supported by the National Scientific Program "Information and Communication Technologies for a Single Digital Market in Science, Education and Security" of the Bulgarian Ministry of Education and Science.

\ifCLASSOPTIONcaptionsoff
  \newpage
\fi



%
\bibliographystyle{IEEEtran}
\bibliography{refs}

%

\end{document}